# Anisotropy of excitation and relaxation of photogenerated Dirac electrons in graphene

*Martin Mittendorff[1,2], Torben Winzer[3], Ermin Malic[3], Andreas Knorr[3], Claire Berger[4,5], Walter A. de Heer[4], Harald Schneider[1], Manfred Helm[1,2], and Stephan Winnerl[1]*

[1] Helmholtz-Zentrum Dresden-Rossendorf, P.O. Box 510119, 01314 Dresden, Germany

[2] Technische Universität Dresden, 01062 Dresden, Germany

[3] Technische Universität Berlin, Hardenbergstraße 36, 10623 Berlin, Germany

[4] Georgia Institute of Technology, Georgia, Atlanta 30332, United States

[5] CNRS – Institut Néel, 38042 Grenoble, France

We investigate the polarization dependence of the carrier excitation and relaxation in epitaxial multilayer graphene. Degenerate pump-probe experiments with a temporal resolution of 30 fs are performed for different rotation angles of the pump-pulse polarization with respect to the polarization of the probe pulse. A pronounced dependence of the pump-induced transmission on this angle is found. It reflects a strong anisotropy of the pump-induced occupation of photogenerated carriers in momentum space even though the band structure is isotropic. Within 150 fs after excitation an isotropic carrier distribution is established. Our observations imply the predominant role of collinear scattering preserving the initially optically generated anisotropy in the carrier distribution. The experiments are well described by microscopic time-, momentum, and angle-resolved modelling, which allows us to unambiguously identify non-collinear carrier-phonon scattering to be the main relaxation mechanism giving rise to an isotropic distribution in the first hundred fs after optical excitation.

The peculiar band structure of graphene gives rise to a number of fascinating transport phenomena and optical effects [1,2]. Optical properties and carrier dynamics have been investigated on ultrashort timescales by pump-probe experiments involving photon energies from a few meV up to 5 eV [3-11]. Highest temporal resolution (below 10 fs) was reported for photon energies of 1.6 eV [11]. For photon energies above 2 eV a deviation from linear dispersion and triangular warping of the band structure become important, while for smaller photon energies the band structure is described in good approximation by an isotropic Dirac cone [12]. However, it has been shown theoretically that the polarization direction of linearly polarized radiation breaks the symmetry of the system [13-15]. The resulting distribution of photogenerated carriers exhibits maxima in the momentum directions perpendicular to the polarization direction and nodes in the directions parallel to the polarization direction. Such an anisotropic distribution of carriers generated by interband excitation is also expected for other materials. However, the complex structure of the valence bands in many materials results in an almost isotropic net distribution without nodes [16]. In graphene, on the other hand, the electron-hole symmetry makes the material ideally suited for studying the anisotropy of carrier excitation. Nevertheless this effect has escaped experimental observation in pump-probe experiments so far, despite the large number of studies. Concerning the relaxation dynamics subsequent to optical excitation carrier-carrier and carrier-phonon contributions have been studied experimentally and theoretically [3-10, 14,15]. For carrier-carrier scattering it has been debated controversially how this process, which can satisfy energy and momentum conservation for two-particle interactions only in case of collinear scattering, can contribute to the carrier dynamics on the two-dimensional manifold of the band structure [17,18,19]. It has been predicted that interband carrier-carrier scattering, often called impact ionization or inverse Auger scattering, can result in carrier multiplication [15,20]. This interesting effect has recently been observed experimentally [8,19,21].

Our polarization-sensitive pump-probe experiments provide direct evidence for an anisotropic occupation. The carriers are generated by linearly polarized radiation of photon energy 1.55 eV, i.e. within the isotropic Dirac-cone region. An isotropic distribution is established within the first 150 fs after optical excitation. It is reached by non-collinear scattering, mainly mediated by optical phonons as unambiguously shown by our microscopic calculations. The faster decay component of the induced transmission observed for pump and probe beams with parallel polarization points towards strong collinear scattering preserving the initially generated anisotropic carrier distribution.

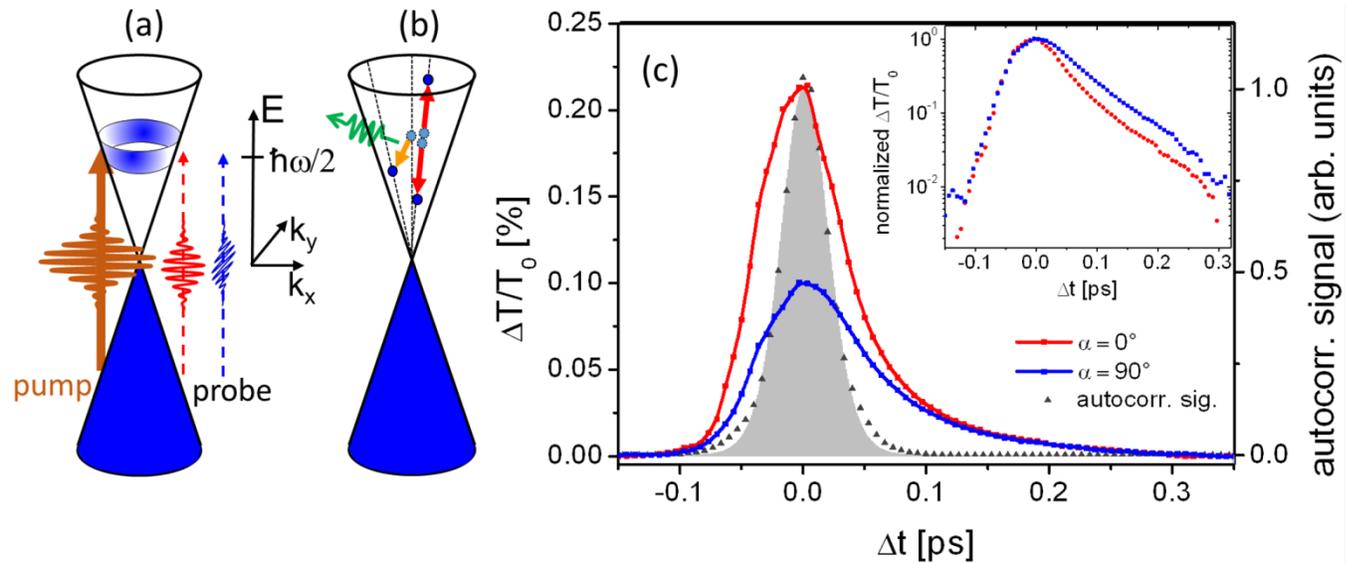

**Figure 1.** (a): Band structure of graphene with anisotropic conduction band occupation induced by a linearly polarized pump beam. (b): Illustration of collinear (red arrows) scattering and scattering across the Dirac cone (orange arrow). Collinear intraband carrier-carrier scattering preserves both energy and momentum while scattering across the Dirac cone involves phonons. (c): Pump-Probe signals for parallel polarization configuration ($\alpha = 0°$) and perpendicular polarization configuration ($\alpha = 90°$). The dots/squares are experimental data, the lines guides for the eye. The black triangles correspond to the measured autocorrelation signal, the grey shaded area is a fit to the autocorrelation data. For better comparison the curves are shifted along the horizontal axis so that the maxima occur at $\Delta t=0$. In the inset the same pump-probe data as in the main figure are depicted, however normalized and plotted on a logarithmic scale.

A Ti:sapphire laser served as a source for 30-fs pulses with a photon energy of 1.55 eV and a spectral width (full width at half maximum FWHM) of 0.20 eV (repetition rate 78 MHz, pulse energy ~3 nJ). The polarization of the pump pulses was varied by rotating a half-wave plate. The probe-beam path contained a second half-wave plate with fixed orientation, hence both pump and probe pulses acquire equal amounts of dispersion. The stretching of the pulses by dispersion was partly compensated by introducing a down-chirp in front of the setup. An off-axis parabolic mirror focused both the pump and probe beam onto the sample. The applied pump fluence was 4 $\mu$J/cm$^2$. The pulse duration at the sample position was characterized by quadratic autocorrelation with a beta barium borate (BBO) crystal. The measured autocorrelation signal can be well described by a self-convolution of a sech$^2$-function with FWHM of 30 fs (cf. Fig. 1). The pump-induced transmission change of graphene was recorded via lock-in detection based on modulation of the pump beam by a chopper. Scattered pump radiation was suppressed by spatial filtering in front of the photodetector. The experiments were performed at room temperature on two samples of epitaxial multilayer graphene (~50 layers and ~70 layers, respectively) grown by thermal decomposition of SiC on the C-terminated face of SiC [22]. Both samples showed similar effects, the data presented in this article stem from a sample with ~50 graphene layers. The graphene-like nature of the layers was confirmed by Raman spectroscopy [23] and magneto-spectroscopy [24].

In Fig. 1 the induced transmissions for parallel and perpendicular polarization orientation of pump and probe beam are presented. For the case of parallel polarization the signal amplitude is slightly more than two times larger than for the case of perpendicular polarization. For time delays beyond 150 fs no differences in the signals are found indicating that an isotropic carrier distribution is reached on this timescale (cf. Fig. 1c). For the configuration with perpendicular polarizations a single-exponential decay with a time constant of 65 ± 5 fs was observed (cf. inset of Fig. 1c). A similar decay occurs in case of parallel polarization for delay times beyond 50 fs. For shorter delay times, however, a very fast initial decay is revealed (cf. inset of Fig. 1c). The basically similar rise and fall time of the signal for parallel polarizations in the time interval [-50 fs, 50 fs] indicates that the fast decay is characterized by a time constant smaller than the pulse duration of 30 fs. The fact that this component is much faster than the timescale for establishing an isotropic distribution points towards pronounced collinear scattering preserving the initially anisotropic carrier distribution. In Fig. 1b a collinear carrier-carrier scattering process along a line on the Dirac cone is illustrated.

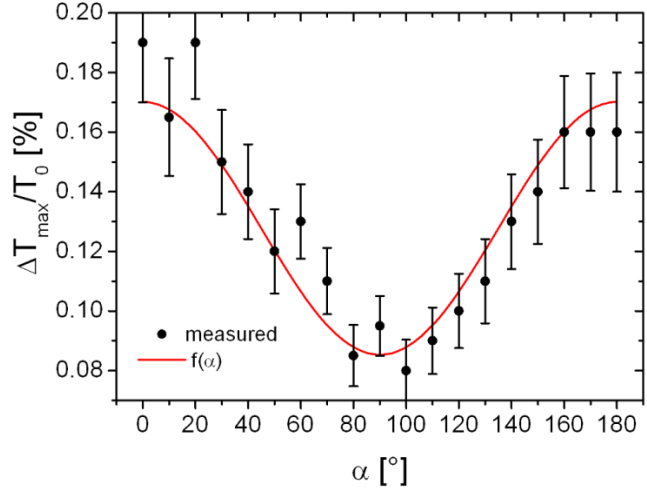

**Figure 2.** Dependence of the amplitude of the pump-probe signal on the angle α between pump and probe-beam polarizations. The fit function f(α) is described in the main text.

To study the anisotropy of the optical excitation in more detail the angle α between pump and probe polarization was varied in steps of 10°. The measured amplitude of the signal follows a function f(α) = $A_0$ + $A_1$ cos$^2$(α) with $A_0 \approx A_1$ (cf. Fig. 2). To verify that the anisotropy purely stems from the optical excitation the sample was rotated and similar experiments were performed. No dependence of the pump-probe signals on sample orientation was found (not shown). The distribution of photogenerated carriers is determined by the absolute square of the matrix element for the optical excitation, which is proportional to |sinΦ$_k$|$^2$ [13,14]. The angle Φ$_k$ is the angle between the momentum $\hbar \bm{k}$ of the photogenerated electrons in the conduction band and the polarization $\bm{e}$ is of the radiation. Pauli blocking caused by pump-excited carriers is responsible for the induced transmission $\Delta T/T_0$. Taking into account the carrier distributions generated by the pump and probe beam, respectively, and integrating over Φ$_k$ results in $\Delta T/T_0 \sim 1 + 2\cos^2(\alpha)$. This dependence is characterized by a maximum-to-minimum ratio of 3:1. The somewhat lower value of ~2:1 found in the experiment is a consequence of the finite temporal resolution of the experiment where carrier scattering occurs already during the excitation pulse. Note that even collinear scattering, which preserves the anisotropy of the carrier distribution, reduces the measured maximum-to-minimum ratio, as it scatters carriers out of the probed energy range.

Finally we compare the experimental results with microscopic calculations based on the density matrix formalism. The carrier dynamics in the valence and conduction band is resolved in momentum, angle and time [14,15,25]. Momentum-dependent rates for carrier-carrier and carrier-phonon scattering are microscopically obtained by applying the second-order Born-Markov approximation. The calculations are performed for excitation pulses with photon energy, pulse duration and pump fluence like in the experiment. In Fig. 3a the carrier populations in the conduction band generated at $\Phi_k = 0°$ and $\Phi_k = 90°$, respectively, are compared. The initial population at $\Phi_k = 0°$ is zero representing a node of the distribution. The population builds up due to non-collinear scattering. After 30 fs the population decreases slowly indicating that scattering out of this phase-space region dominates over in-scattering. In contrast the population at $\Phi_k = 90°$ is initially strong and decreases successively. The calculated induced transmission is depicted in Fig. 3b. In very good qualitative and quantitative agreement with the experiment an amplitude ratio slightly above 2:1 is found for parallel and perpendicular polarization configuration. Also the subsequent decay dynamics agrees very well with the experiment (cf. Fig. 1c and 3b), considering that the modelling does not contain adjustable fitting parameters.

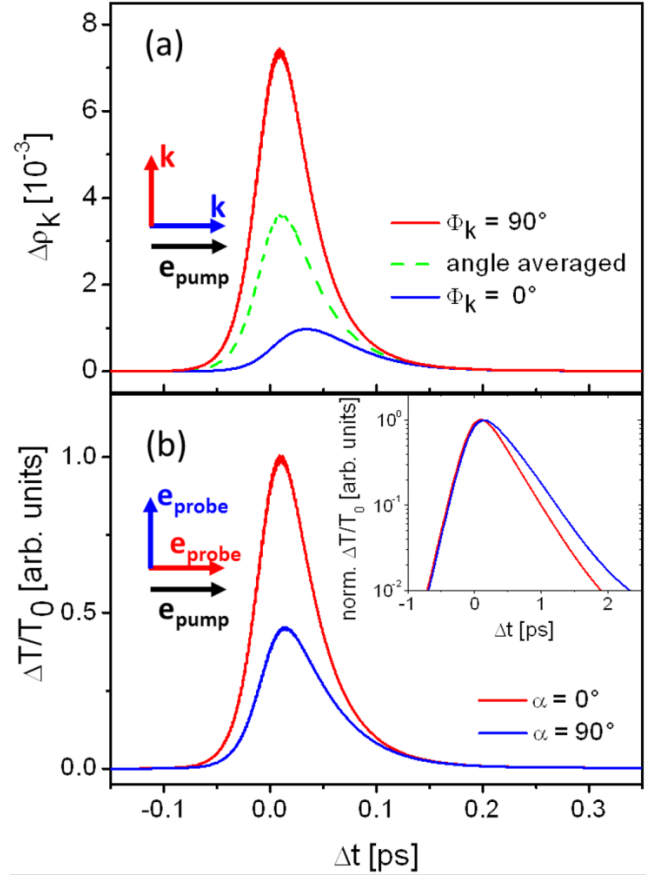

**Figure 3.** (a): Calculated pump-induced carrier occupation in the k-space regions corresponding to $\Phi_k = 0°$, $\Phi_k = 90°$ and averaged over all values of $\Phi_k$. (b): Calculated pump-induced transmission for parallel polarization configuration ($\alpha = 0°$) and perpendicular polarization configuration ($\alpha = 90°$). In the inset the same data as in the main figure are depicted, however normalized and plotted on a logarithmic scale.

Our calculations clearly show that intraband carrier-carrier scattering is the predominant collinear-scattering channel preserving the initial anisotropy. The main process resulting in an isotropic distribution, on the other hand, is found to be the scattering across the Dirac cone via optical phonons. A detailed analysis of the role of the different elementary processes in the relaxation dynamics is provided in Ref. [15].

When comparing the induced transmission for perpendicular polarizations with the population at $\Phi_k = 0$ a pronounced difference in the rise time stands out (cf. Fig. 3a and Fig. 3b). As discussed above, the population at $\Phi_k = 0$ builds up by scattering. In contrast, the induced transmission rises with a rate determined by the pump-pulse duration similarly to the induced transmission for the parallel polarization configuration. This similar rise is also found in the experiment. In case of perpendicular polarizations the rise is caused by the overlap of populations generated by pump and probe beam, in particular in the phase-space regions around $\Phi_k = 45°, 135°, 225°$ and $315°$.

We note that signatures of the optical anisotropy have been predicted [26,27] and observed [28] for photocurrents in electrically contacted graphene layers. In this case an additional symmetry breaking is required as the net momentum photocurrent is zero, as positive and negative momentum contributions cancel each other out. A bias field provides a preferred direction for the photocurrent. The anisotropy of measured photocurrents exhibits a maximum-minimum ratio of ~6:5. The lower value as compared to our experiment results possibly from scattering, which contributes considerably stronger in the transport experiments.

Our findings have several interesting implications: Firstly, pumping and probing with orthogonal polarization directions is often applied by experimentalists for purely technical reasons, namely in order to minimize the effects of scattered pump radiation. This technique has also been employed in a number of studies on graphene [3,7,11]. Our results show that this has no influence on the observed dynamics on timescales beyond 150 fs. On shorter timescales, however, one has to be aware that different relaxation dynamics are probed by the two polarization schemes. Secondly, the anisotropy in excitation and saturation may be exploited in all-optical switches that react differently to pulses of different polarization direction.

Furthermore the observation of the pronounced anisotropy in optical excitation will pave the way to experiments with more complex excitation conditions. In this respect ω-2ω-schemes have been proposed in theoretical studies [29], where the distribution of excited carriers in *k*-space can be controlled by varying the phase between the ω and 2ω pulse giving rise to pure optical currents [30].

In conclusion, a pronounced anisotropy in both the optical excitation and the subsequent ultrafast relaxation dynamics has been revealed for electrons within the isotropic Dirac cone in graphene. The first effect is directly related to the anisotropy of the matrix elements for excitation with linearly polarized radiation. After 150 fs an isotropic distribution is established. Furthermore, evidence is found for Coulomb-induced collinear scattering preserving the anisotropy on much shorter timescales. The knowledge of orientational relaxation of carrier momentum distributions as obtained in this work is highly relevant for better understanding of ballistic photocurrents in graphene and for device applications based on anisotropy effects in this material.

AUTHOR INFORMATION

**Corresponding Author**

* m.mittendorff@hzdr.de

ACKNOWLEDGMENT

We acknowledge the financial support by the German Research Foundation (DFG) through SPP 1459. E. M. is grateful to the Einstein Foundation Berlin.

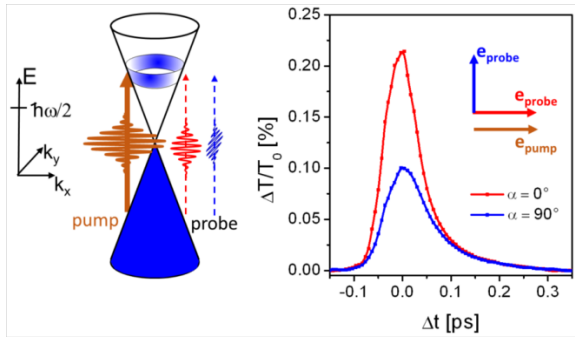

For Table of Contents only